\documentclass{emulateapj}

\newcommand{\simge}{\mbox{$\stackrel{>}{_{\sim}}$}}

\newcommand\msun{\hbox{\,M$_\odot$}}

\newcommand\lsun{\hbox{\,L$_\odot$}}

\slugcomment{Accepted for publication in ApJ Letters}

\shorttitle{The Circumbinary Disk of v892 Tau}
\shortauthors{Monnier et al.}

\begin{document}


\title{Discovery of a Circumbinary Disk around Herbig Ae/Be system v892~Tau}


\author{J.~D.~Monnier\altaffilmark{1}, 
A. Tannirkulam\altaffilmark{1}, 
P. G. Tuthill\altaffilmark{2},
M. Ireland\altaffilmark{2}, R. Cohen\altaffilmark{3},
W. C. Danchi\altaffilmark{4},
F. Baron\altaffilmark{5}
}

\altaffiltext{1}{monnier@umich.edu: University of Michigan Astronomy Department, 941 Dennison Bldg, Ann Arbor, MI 48109-1090, USA.}
\altaffiltext{2}{University of Sydney}
\altaffiltext{3}{W. M. Keck Observatory}
\altaffiltext{4}{NASA-GSFC}
\altaffiltext{5}{University of Cambridge}
\email{JDM: monnier@umich.edu}


\begin{abstract}
  We report the discovery of a circumbinary disk around the Herbig
  Ae/Be system v892~Tau.  Our detailed mid-infrared images were made
  using segment-tilting interferometry on the Keck-1 Telescope and
  reveal an asymmetric disk inclined at $\sim$60$\arcdeg$ with an
  inner hole diameter of 250~mas (35~AU), approximately 5$\times$
  larger than the apparent separation of the binary components.  In
  addition, we report a new measurement along the binary orbit using
  near-infrared Keck aperture masking, allowing a crude estimate of
  orbital parameters and the system mass for the first time.  The size
  of the inner hole appears to be consistent with the minimum size
  prediction from tidal truncation theory, bearing a resemblance to
  the recently unmasked binary CoKu Tau/4.  Our results have motivated a
  re-analysis of the system spectral energy distribution, concluding
  the luminosity of this system has been severely underestimated.
  With further study and monitoring, v892~Tau should prove a
  powerful testing ground for both predictions of dynamical models for
  disk-star interactions in young systems with gas-rich disks and for
  calibrations of pre-main-sequence tracks for intermediate-mass stars.


\end{abstract}





\section{Introduction}

v892~Tau (Elias-1, Elias 3-1) is a young stellar object in the
Taurus-Aurigae star forming region \citep[$d=140~pc$,][]{kdh1994}.
Its visible spectrum is classified as spectral type B8V
\citep{hernandez2004},
making v892~Tau one of the few Herbig Ae/Be
stars in Taurus (in addition to AB~Aur, MWC 480).  The spectral energy
distribution (SED) of v892~Tau is dominated by bright thermal emission
in the mid-infrared \citep{hillenbrand1992}, with relatively large
line-of-sight extinction in the visible \citep[one estimate is A$_{\rm
  V}\sim5.9$,][]{kh1995}.  Despite being one of the closest Herbig
stars and after decades of multi-wavelength observations, the nature
of v892~Tau is still uncertain.

The SED suggests v892~Tau is either a young embedded Class I source
\citep{lada1987} still surrounded by its nascent envelope or a more
evolved, Class II object (i.e, Herbig Ae/Be star) seen through its
edge-on disk.  
Spatially-resolved imaging can easily distinguish between
these two scenarios, and early speckle
interferometry \citep{kataza1991,haas1997,leinert2001} suggested the
presence of an extended and elongated nebula in the near-infrared that
could be due to scattering in bipolar lobes.

The high-resolution speckle imaging of \citet{smith2005} clearly
resolved the K band ($\lambda=2.2\mu$m) emission to be coming from
two unresolved stars in v892~Tau with little or no sign of a nebula or 
extended emission.  The two stars of roughly equal brightness had an 
apparent separation of 55~milliarcseconds (7.7~AU) and some evidence of orbital 
motion was seen between epochs separated by 7 years.  

Although this binary should have a dramatic effect on the infrared
emission, carving out a large hole in the circumbinary disk, 
recent analysis of the SED of v892~Tau including ISO data
\citep{acke2004} uncovered no distinct signature of the underlying
binary\footnote{\citet{acke2004} did note strong 11.2$\mu$m PAH emission and
  anomalous infrared colors.}.  Other workers \citep{liu2005,liu2007} made use
of the technique of nulling interferometry in the mid-IR
($\lambda=10.3\mu$m) to marginally resolve v892~Tau (FWHM $\sim$20~AU)
along PA~164$\arcdeg$ consistent with normal (single-star) disk emission.

Here, we report new mid-iIR imaging of the v892~Tau system which
resolves these mysteries, discovering very extended and resolved
emission that we interpret to be coming from a circumbinary disk.
We also confirm the presence of the binary at K
band and our new data allow first crude estimates of the orbital elements. 
Lastly, we report a new SED analysis which 
provides an improved determination of the system luminosity and our 
viewing geometry.

\section{Observations}
\label{section:observations}

We observed v892~Tau using the Keck-1 telescope as part of two
separate experiments, the Keck segment-tilting experiment and the Keck
aperture masking experiment.  Here, we briefly describe these
experiments and give pertinent observing details.

%
v892~Tau was observed on UT2004Aug31, UT 2004Sep01, and UT2005Feb19 at
the Keck-1 telescope using the 10.7$\mu$m filter
($\lambda_0=10.7\mu$m, $\Delta\lambda=$1.55$\mu$m)
 of the Long Wavelength Spectrometer
\citep[LWS,][]{campbell2004} just before this instrument was
decommissioned in 2005.  In order to optimize our
calibration against changes in seeing \citep[e.g.,][]{tuthill2000b},
we used the Keck's segmented primary mirror in a novel ``segment-tilting''
mode, whereby we controlled the tilts and pistons of the individual
mirror segments to create a set of 4 independent and non-redundant
interference patterns (using 6 segments each) on the LWS detector
focal plane.  Furthermore, short exposures were used ($t_{\rm
  int}=90$~ms$<t_0$, where 
  $t_0$ is the atmospheric coherence time $\sim$250~ms in the mid-iR) to effectively freeze the atmospheric turbulence.
For calibration we interleaved target observations with calibrators
$\alpha$~Tau\footnote{All calibrators were assumed unresolved except
  $\alpha$~Tau where we adopted a uniform disk diameter of 19.8~mas
  \citep{perrin1998}.} , $\iota$~Aur, and $\alpha$~Cet.  Details on
this experiment and the implementation at Keck were first introduced
by \citet{monnier_aas2004} and more information can be found in
recently published science papers
\citep{weiner2006,ireland2007,rajagopal2007}.



\begin{figure}[hbt]
\begin{center}
\includegraphics[angle=90,width=7in]{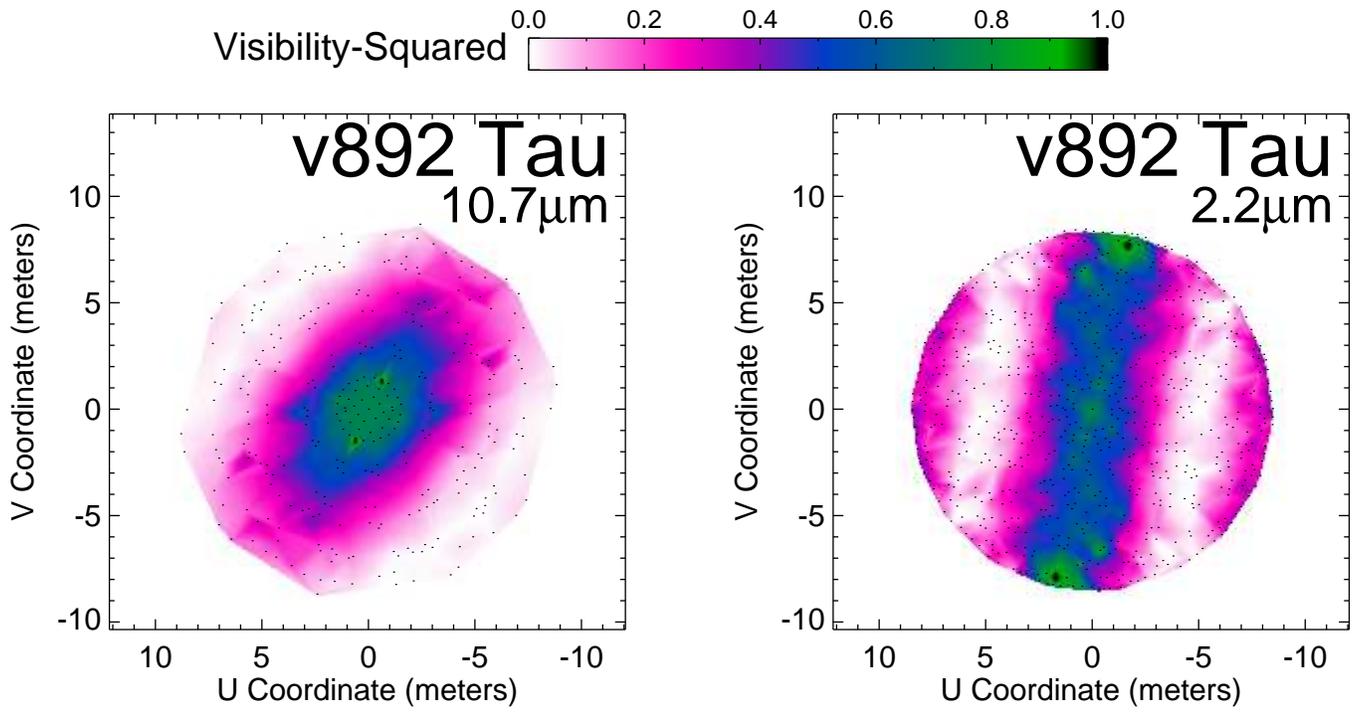}
\figcaption{\footnotesize This figure shows visibility
  data (color scale) for v892~Tau and the specific Fourier coverage (dots).
  The left panel shows results for 10.7$\mu$m using the Keck segment tilting method while the right panel applies to the 2.2$\mu$m Keck aperture masking data.
\label{fig_vis2}}
\end{center}
\end{figure}

The combined Fourier coverage of the four patterns over the three
nights (total of 6 independent pointings) can be found in
Figure~\ref{fig_vis2} along with the visibility results.  The
visibility-squared data and the closure phases were compiled and saved
using the OIFITS data format \citep{pauls2005} and are available upon
request.  Figure~\ref{fig_vis2} shows the
clear sign of highly-elongated disk emission.

We also
obtained diffraction-limited observations of v892~Tau in the
near-IR on UT2004Sep04 using the K-band filter
($\lambda=$2.21$\mu$m, $\Delta\lambda=$0.43$\mu$m) of the NIRC camera
\citep{matthews96} in conjunction with an annulus aperture mask placed
in front of the secondary mirror.  This observing mode was extensively
utilized between 1997 and 2005 \citep[see most recent science
papers,][]{monnier2007a, tuthill2008a} and details of
the experimental design and performance 
can be found in \citet{tuthill2000b}.  We interleaved
observations of the target with the unresolved calibrator 54~Per.  The
Fourier coverage and visibility-squared results are shown in
Figure~\ref{fig_vis2}, where the the binary nature is clearly revealed.

\section{Analysis}

\subsection{Image Reconstructions}
\label{section_imaging}
\begin{figure}[hbt]
\begin{center}
\includegraphics[angle=90,width=7in]{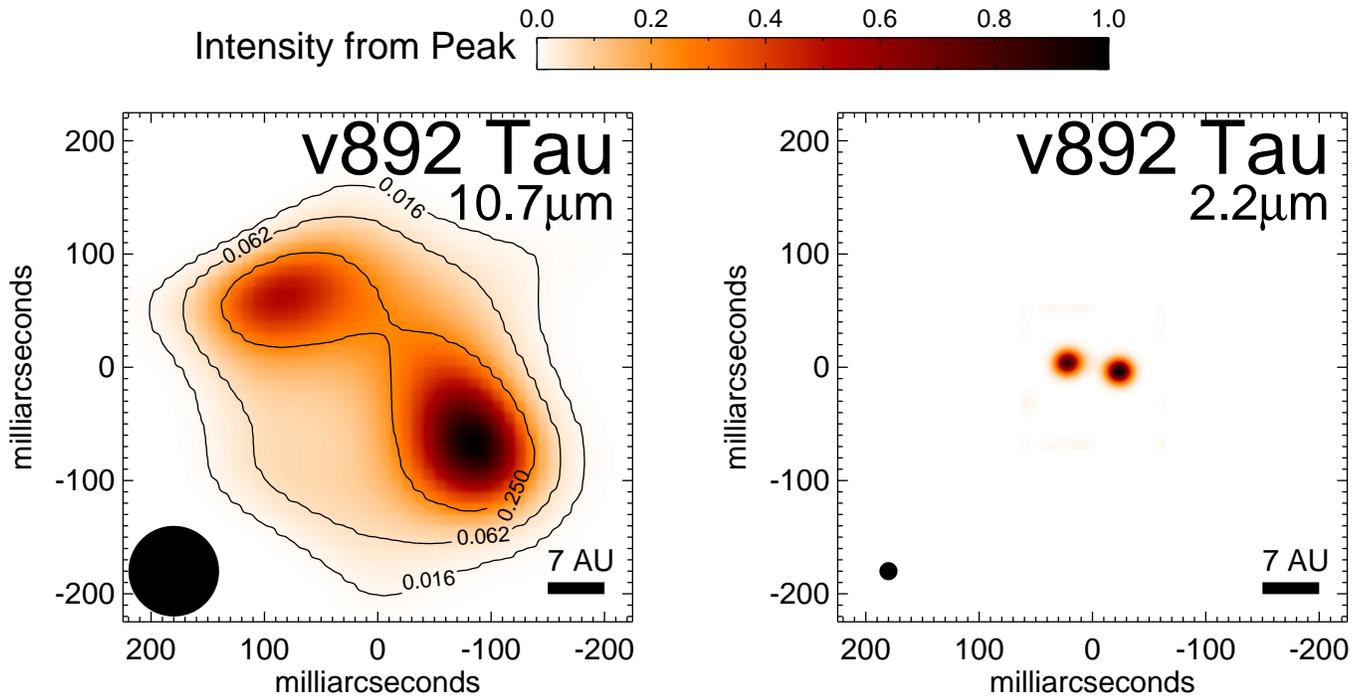}
\figcaption{\footnotesize This figure shows the image reconstructions
  of v892~Tau at 10.7$\mu$m (left panel) and 2.2$\mu$m (right panel)
  using the BSMEM software. At the bottom-left of each panel, we have
  included an estimate of the resolution of each image, corresponding
  here to 80~mas at 10.7$\mu$m and 16~mas at 2.2$\mu$m.  Contour
  levels are shown for the extended emission in the mid-IR
  spaced logarythmically in factors of 4: 25\%, 6.3\%, and 1.6\% of
  the peak. The scale bar applies for a distance of 140~pc.  
  For orientation, East points left and North points up.
\label{fig_images}}
\end{center}
\end{figure}

We used the BSMEM image reconstruction software
\citep{buscher1994,lawson2004,lawson2006} for aperture synthesis
imaging.  Figure~\ref{fig_images} shows the reconstructed images for
the mid-IR and near-IR data.  In order to confirm the
asymmetric features in the mid-IR image were not due to artifacts in
the BSMEM algorithm, we also used the independent image reconstruction
code MACIM \citep{macim} and found good consistency between the basic
morphology, size scale, and level of emission asymmetry.  As an
additional data quality check, separate images were made for each
independent night of data and the resulting images all closely
resembled the result shown in Figure~\ref{fig_images}.  Note that we do not know
the relative position of the near-IR image compared to  mid-IR image
and have presented each image centered in its corresponding
frame.

\subsubsection{Mid-infrared: Elongated disk structure}

The mid-IR image shows elongated emission, approximately 320~mas
by 180~mas in full extent, elongated along PA~50$\arcdeg$.  The two
bright lobes in the image are separated by 210~mas (30~AU) with the
south-west side being significantly brighter.  A two-dimensional
Gaussian fit to the visibility data gives a FWHM 244$\pm$6~mas
$\times$ 123$\pm$9~mas along PA 49$\pm$1$\arcdeg$.  For a simplistic
``flat disk'' model, this 2-to-1 ratio elongation suggests 
we are viewing this disk oriented at $\simge$60$\arcdeg$ inclination.

A more suitable model for a circumbinary or transitional disk
would be a tilted and asymmetric ring of emission that
is nearly unresolved along the minor axis \citep[for example, see dramatic
case of HR 4796;][]{koerner1998,schneider1999}.  Following the 
parameterization of \citet{monnier2006}, we have fitted a ``skewed
asymmetric ring model'' to our interferometry data here to better estimate the
inner hole diameter.  
The best fitting model had an inner-hole diameter of 
247$\times$121~mas (35$\times$17 AU) along PA 53$\arcdeg$, with a 40\%
skew along PA 284$\arcdeg$ (the thickness of the ring followed a
gaussian profile with a FWHM of 25\% ring radius).  For purposes of
\S\ref{discussion}, we will assume the central hole in the
circumbinary disk has a radius of 17.5~AU.


Importantly, we want to emphasize that the scale of the emission is
more than 4$\times$ larger than the separation of the binary at the
heart of this system, and the elongation is along a
distinctly-different position angle.  Since our mid-IR data in
2004 were taken within one week of the near-IR data, we can
clearly prove that the elongated mid-IR emission comes from a
distinct and much larger component than the near-IR.  We discuss
the relationship between the circumbinary disk and the underlying
binary orbit in \S\ref{discussion}.

At first glance, it may seem surprising that \citet{liu2005} did not more
clearly resolve this large circumbinary disk using their nulling technique,
finding a FWHM of only 20~AU. However, this nulling measurement was
performed only for PA~164$\arcdeg$, which is along the narrow dimension
of the elongated emission we detected.  Another possible explanation
for the smaller size is that the bandpass filter used here has more
contribution from 11.2$\mu$m PAH emission, which could be more extended
than the mid-IR continuum.

\subsubsection{Near-infrared: Binary star}
Figure~\ref{fig_images} shows the BSMEM image of v892~Tau at
2.2$\mu$m.  We confirm the binary nature of this target as first
reported by \citet{smith2005}. We also set a limit of $<$10\% of the
emission possibly coming from any sort of extended or ``halo''
component \citep[e.g.,][]{kataza1991,haas1997, leinert2001}.  In order
to extract the separation and position angle of this binary, we have
used both image fitting and direct fitting to the interferometric
observables to yield the following result: $\rho=$44.2$\pm$1.0~mas,
$\theta=$79.9$\pm$1.0$\arcdeg$, Flux Ratio 1.15$\pm$0.04 (west
component brighter than east component).

\subsection{Binary orbit}
\label{orbit}
\begin{figure}[hbt]
\begin{center}
\includegraphics[angle=90,height=2.5in]{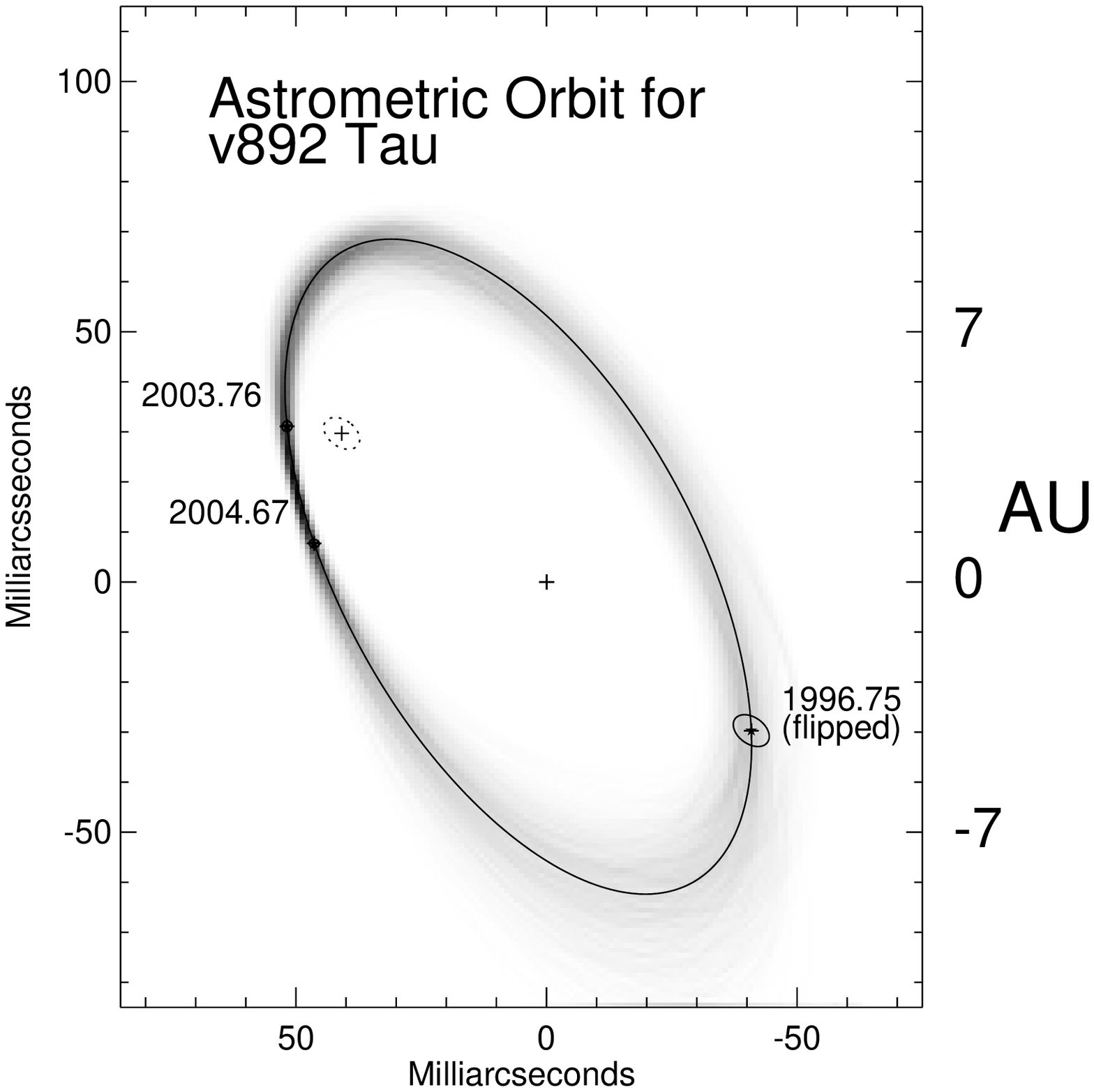}
\includegraphics[angle=90,height=2.5in]{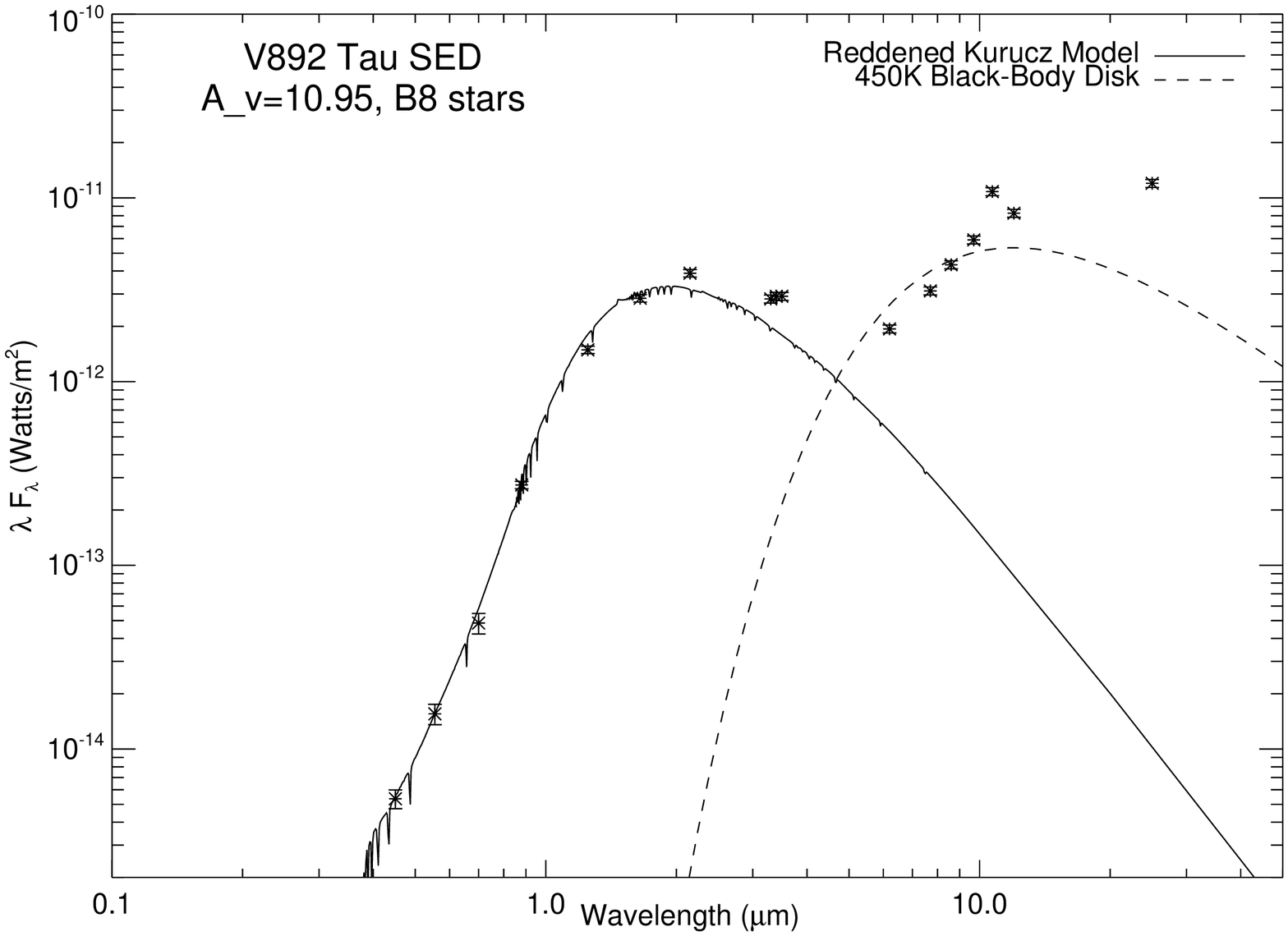}
\figcaption{\footnotesize {\em (left panel)} This figure shows the
  separation and position angle measurements for v892~Tau from
  \citet{smith2005} and this work.  We have plotted the best-fit model
for a system mass of 5.5~$\msun$ (period 13.8 yrs, eccentricity 0.12) with the solid curve.
  Note that we had to flip the position angle of the 1996 epoch
  measurement in order to find a realistic orbit  (see \S\ref{orbit} for full description of our
  procedure and justification). {\em (right panel)} This figure shows
  the spectral energy distribution for v892~Tau 
  \citep[photometry from][and our own JHK photomery from MDM Observatory]{strom1994,acke2004,herbst1999}
and our proposed
  decomposition of star plus disk. We have plotted the contribution
  from the reddened ($A_V=10.95, R_V=6.4$) stellar photospheres
  (2$\times$B8V, system luminosity of 400$\lsun$) and the contribution
  from the ``warm inner wall'' of the circumbinary disk with
  temperature T$\sim$450K.
\label{fig_orbit}}
\end{center}
\end{figure}

We combined our new binary measurement (at 2004.67) with the 1996.75 and 2003.76
measurements from \citet{smith2005} and these data are plotted in
Figure~\ref{fig_orbit}.  The small proper motion ($\Delta\sim$10~mas$=1.4$~AU)
observed by \citet{smith2005} over 7~years is problematic for this
object since this would require a long period orbit ($\simge$100~yrs) 
inconsistent with
the spectral types and luminosities of the binary stars themselves.

Alternatively, the binary period could be approximately 7~years, meaning the
stars had gone nearly exactly once around their orbit between epochs.  Although
the timing is suspicious, the scenario is plausible except this
would require a system mass $>$20$\msun$, much too high for the B8
spectral type and system luminosity.

The scenario we favor is that infrared variability affected the
brightness ratio between 1996 and 2003, confusing the assignment of
``primary'' and ``secondary'' star. This variation could be intrinsic
or be due to varying line-of-sight obscuration through this asymmetric
disk.  \citet{smith2005} did report a change in flux ratio, finding a
brightness ratio of 1.0 for 2003.76 which implies that their reported
position angle had a $\pm$180$\arcdeg$ ambiguity.

In our 2004.67 data, we definitively detect the south-west component
as 15\% brighter, consistent with the PA assignment of the published
2003.76 measurement.  By flipping the position angle of the earlier
1996 measurements, we find a robust family of orbits with periods of
$\sim$14~years, compatible with the measured binary separations and
the expected system mass.  Note that future orbital refinement will
lead to precise stellar masses for these intermediate-mass,
pre-main-sequence stars \citep[see also MWC~361A,][]{monnier2006} -- a
rare opportunity to advance the calibration of pre-main-sequence
tracks for this mass range.

To estimate the orbital parameters for v892~Tau, we used Monte Carlo sampling
of the measured stellar separations along with a system mass constraint of
5.5$\pm$0.5$\msun$ expected for two stars of spectral type B8V \citep{palla1993}.
Figure~\ref{fig_orbit} shows the main results of our orbital study: $P=13.8\pm1.5$~yrs, $a=72.4\pm6.3$~mas (10~AU), $e=0.12\pm0.05$,
$i=60.6\pm3.8\arcdeg$, $\omega=233\pm42\arcdeg$, $\Omega=28\pm5\arcdeg$, 
$T_O = 55480\pm900$~MJD. 

\section{Discussion}
\label{discussion}
We interpret the mid-IR emission of v892~Tau as a circumbinary
disk based on several arguments. The mid-IR emitting region is much
larger than the separation between the binary components, ruling out
models which have all the mid-IR emission coming from disks around the
individual stars. In theoretical models of tidal truncation
\citep{artymowicz1994}, the circumbinary disk has an inner hole
approximately 1.8-2.6$\times$ larger than the semi-major axis of the
binary system (for eccentricity between 0 and 0.25). For comparison,
our best estimates of the hole radius and orbital semi-major axis are
17.5 AU and 10~AU respectively -- giving a ratio of 1.75 which is
close to the theoretical expectation.  Additionally, the position
angle and derived inclination angle of the mid-IR emission 
(PA$\sim$50$\arcdeg$, $i\sim$60$\arcdeg$) 
is similar to
those derived from the orbit (PA$\sim$28$\arcdeg$, $i\sim$61$\arcdeg$) 
and the mild
asymmetry in the mid-IR emission suggests dynamical interactions
between an eccentric binary and the surrounding disk (via resonances or disk warping).

We further test the circumbinary disk hypothesis by analyzing the
spectral energy distribution (SED). The SED shows a large mid-IR bump
similar to that seen in ``transitional'' disks
\citep{calvet2002,espaillat2007}.  By reddening template Kurucz spectra for
B8V stars, we can fit the near-IR emission with mostly photospheric
light (some near-IR emission from hot dust is allowed but not well
constrained) assuming A$_V$ = 10.95 yielding a combined stellar
luminosity of 400~L$_{\odot}$, reasonable for two B8V stars.  Our new
proposed SED decomposition for v892~Tau can be found in
Figure~\ref{fig_orbit}, showing that 
the rising mid-IR flux can be roughly characterized as a
T$\sim450$~K blackbody, consistent with emission from
the warm inner wall of the circumbinary disk.
 
Using equation (1) in \citet{isella2006}, the predicted radius for the
warm inner wall for the above stellar luminosity (assuming 0.25 $\mu$m spherical
silicate grains) is 18~AU -- consistent with our derived inner hole radius
of 17.5~AU based on imaging.  Furthermore, we find that the scale
height of this wall at 18 AU is $\sim$ 1.8 AU \citep{ddn2001} which
can produce the observed line-of-sight A$_V \sim$ 11 for the disk
inclination of $\sim$65$^o$. This latter phenomena may not be widely
appreciated -- circumbinary (or transitional) disks have warm
puffed-up inner walls (the rim scale height is a stronger function of
radius than temperature) at large radii that effectively increases the
possibility that central stars will be obscured or at least reddened
as seen by distant observers.  The lack of scattering nebulosity in archival Hubble Space Telescope images
independently suggests that the high line-of-sight A$_V$ is likely from absorption by dust in the outer disk and not infalling envelope material.
While a detailed model is beyond the
scope of this Letter,  a careful study will allow precise
constraints of stellar luminosities, dust properties, and circumbinary
structure and should be straight-forward with today's 3-D Monte Carlo
radiative transfer codes \citep[e.g., TORUS,][]{torus}.

\section{Conclusions}
We have discovered an extensive circumbinary disk around v892~Tau in
the mid-infrared.  We also independently confirm the binary nature of
the underlying stellar system and our new measurement allows us to fit
an astrometric orbit, finding a period $\sim$14~years for system mass of $\sim$6$\msun$.
Our limited orbital phase
coverage and some ambiguity in position angles allow only
crude estimates and we strongly urge continued monitoring of this
system using near-IR speckle, aperture masking, or adaptive optics.

We have proposed a new SED decomposition with a line-of-sight
extinction ($A_V\sim11$) higher than previously thought, implying a
system luminosity $\sim$400$\lsun$.  This result highlights
the fact that circumbinary disks (and transitional disks) have much
larger opening angles since the puffed-up inner wall causes 
enhanced extinction of the central stars even at
inclinations of 60$\arcdeg$.

v892~Tau is another case where high-resolution imaging has motivated a
fundamental shift in our understanding of an individual object.  As
spectral energy distributions are used to discover ``transitional''
disks implicating planet formation, we must be cognizant of the
important role of binarity and the associated circumbinary disks that
can mimic signs of planet formation \citep[e.g.,][]{ireland2008}.
Far from merely being spoilers to planet finders, new circumbinary
disks offer fresh laboratories for studying dynamical interactions
between gas-rich disks and the massive orbiting bodies embedded within
them as well as critical opportunities to calibrate pre-main sequence
tracks.

\acknowledgments {We thank Nuria Calvet and Jesus Hernandez for
  helpful discussions and acknowledge support from NASA KPDA grant
  1267021, NASA Origins grant NNG05GI80G, NSF-AST 0352728, and the
  NASA Michelson Fellowship program (MJI). The data presented herein
  were obtained at the W.M. Keck Observatory, which is operated by
  Caltech, University of California and NASA.  WMKO
  was made possible by the financial support of the W.M. Keck
  Foundation.  The authors wish to recognize and acknowledge the very
  significant cultural role and reverence that the summit of Mauna Kea
  has always had within the indigenous Hawaiian community.  We are
  most fortunate to have the opportunity to conduct observations from
  this mountain.

{\it Facility:} \facility{Keck:I (LWS,NIRC)} \facility{Hiltner (TIFKAM)}
}

\bibliographystyle{apj}
\bibliography{apj-jour,ms}

\begin{thebibliography}{42}
\expandafter\ifx\csname natexlab\endcsname\relax\def\natexlab#1{#1}\fi

\bibitem[{{Acke} \& {van den Ancker}(2004)}]{acke2004}
{Acke}, B. \& {van den Ancker}, M.~E. 2004, \aap, 426, 151

\bibitem[{{Artymowicz} \& {Lubow}(1994)}]{artymowicz1994}
{Artymowicz}, P. \& {Lubow}, S.~H. 1994, \apj, 421, 651

\bibitem[{{Buscher}(1994)}]{buscher1994}
{Buscher}, D.~F. 1994, in IAU Symposium, Vol. 158, Very High Angular Resolution
  Imaging, ed. J.~G. {Robertson} \& W.~J. {Tango}, 91--+

\bibitem[{{Calvet} {et~al.}(2002){Calvet}, {D'Alessio}, {Hartmann}, {Wilner},
  {Walsh}, \& {Sitko}}]{calvet2002}
{Calvet}, N., {D'Alessio}, P., {Hartmann}, L., {Wilner}, D., {Walsh}, A., \&
  {Sitko}, M. 2002, \apj, 568, 1008

\bibitem[{{Campbell} \& {Jones}(2004)}]{campbell2004}
{Campbell}, R.~D. \& {Jones}, B. 2004, Advances in Space Research, 34, 499

\bibitem[{{Dullemond} {et~al.}(2001){Dullemond}, {Dominik}, \&
  {Natta}}]{ddn2001}
{Dullemond}, C.~P., {Dominik}, C., \& {Natta}, A. 2001, \apj, 560, 957

\bibitem[{{Espaillat} {et~al.}(2007){Espaillat}, {Calvet}, {D'Alessio},
  {Hern{\'a}ndez}, {Qi}, {Hartmann}, {Furlan}, \& {Watson}}]{espaillat2007}
Espaillat, C., et~ al.  2007, \apjl, 670, L135

\bibitem[{{Haas} {et~al.}(1997){Haas}, {Leinert}, \& {Richichi}}]{haas1997}
{Haas}, M., {Leinert}, C., \& {Richichi}, A. 1997, \aap, 326, 1076

\bibitem[{{Harries}(2000)}]{torus}
{Harries}, T.~J. 2000, \mnras, 315, 722

\bibitem[{{Herbst} \& {Shevchenko}(1999)}]{herbst1999}
{Herbst}, W. \& {Shevchenko}, V.~S. 1999, \aj, 118, 1043

\bibitem[{{Hern{\'a}ndez} {et~al.}(2004){Hern{\'a}ndez}, {Calvet},
  {Brice{\~n}o}, {Hartmann}, \& {Berlind}}]{hernandez2004}
{Hern{\'a}ndez}, J., {Calvet}, N., {Brice{\~n}o}, C., {Hartmann}, L., \&
  {Berlind}, P. 2004, \aj, 127, 1682

\bibitem[{{Hillenbrand} {et~al.}(1992){Hillenbrand}, {Strom}, {Vrba}, \&
  {Keene}}]{hillenbrand1992}
{Hillenbrand}, L.~A., {Strom}, S.~E., {Vrba}, F.~J., \& {Keene}, J. 1992, \apj,
  397, 613

\bibitem[{{Ireland} \& {Kraus}(2008)}]{ireland2008}
{Ireland}, M.~J. \& {Kraus}, A.~L. 2008, \apjl, 678, L59

\bibitem[{{Ireland} {et~al.}(2006){Ireland}, {Monnier}, \& {Thureau}}]{macim}
{Ireland}, M.~J., {Monnier}, J.~D., \& {Thureau}, N. 2006, 
Proceedings of SPIE, Vol. 6268, pp. 62681T 

\bibitem[{{Ireland} {et~al.}(2007){Ireland}, {Monnier}, {Tuthill}, {Cohen}, {De
  Buizer}, {Packham}, {Ciardi}, {Hayward}, \& {Lloyd}}]{ireland2007}
{Ireland}, M.~J., et~al. 2007, \apj, 662, 651

\bibitem[{{Isella} {et~al.}(2006){Isella}, {Testi}, \& {Natta}}]{isella2006}
{Isella}, A., {Testi}, L., \& {Natta}, A. 2006, \aap, 451, 951

\bibitem[{{Kataza} \& {Maihara}(1991)}]{kataza1991}
{Kataza}, H. \& {Maihara}, T. 1991, \aap, 248, L1

\bibitem[{{Kenyon} {et~al.}(1994){Kenyon}, {Dobrzycka}, \&
  {Hartmann}}]{kdh1994}
{Kenyon}, S.~J., {Dobrzycka}, D., \& {Hartmann}, L. 1994, \aj, 108, 1872

\bibitem[{{Kenyon} \& {Hartmann}(1995)}]{kh1995}
{Kenyon}, S.~J. \& {Hartmann}, L. 1995, \apjs, 101, 117

\bibitem[{{Koerner} {et~al.}(1998){Koerner}, {Ressler}, {Werner}, \&
  {Backman}}]{koerner1998}
{Koerner}, D.~W., {Ressler}, M.~E., {Werner}, M.~W., \& {Backman}, D.~E. 1998,
  \apjl, 503, L83+

\bibitem[{{Lada}(1987)}]{lada1987}
{Lada}, C.~J. 1987, in IAU Symposium, Vol. 115, Star Forming Regions, ed.
  M.~{Peimbert} \& J.~{Jugaku}, 1--17

\bibitem[{{Lawson} {et~al.}(2006){Lawson}, {Cotton}, {Hummel}, {Baron},
  {Young}, {Kraus}, {Hofmann}, {Weigelt}, {Ireland}, {Monnier}, {Thi{\'e}baut},
  {Rengaswamy}, \& {Chesneau}}]{lawson2006}
{Lawson}, P.~R., et~al. 2006, Proceedings of SPIE, Vol. 6268, 
  pp. 62681U

\bibitem[{{Lawson} {et~al.}(2004){Lawson}, {Cotton}, {Hummel}, {Monnier},
  {Zhao}, {Young}, {Thorsteinsson}, {Meimon}, {Mugnier}, {Le Besnerais},
  {Thiebaut}, \& {Tuthill}}]{lawson2004}
{Lawson}, P.~R., et~al.
2004, Proceedings of SPIE, Vol. 5491.  p.886

\bibitem[{{Leinert} {et~al.}(2001){Leinert}, {Haas}, {{\'A}brah{\'a}m}, \&
  {Richichi}}]{leinert2001}
{Leinert}, C., {Haas}, M., {{\'A}brah{\'a}m}, P., \& {Richichi}, A. 2001, \aap,
  375, 927

\bibitem[{{Liu} {et~al.}(2005){Liu}, {Hinz}, {Hoffmann}, {Brusa}, {Miller}, \&
  {Kenworthy}}]{liu2005}
{Liu}, W.~M., et~al.  2005, \apjl, 618, L133

\bibitem[{{Liu} {et~al.}(2007){Liu}, {Hinz}, {Meyer}, {Mamajek}, {Hoffmann},
  {Brusa}, {Miller}, \& {Kenworthy}}]{liu2007}
{Liu}, W.~M., et~al.  2007, \apj, 658, 1164

\bibitem[{{Matthews} {et~al.}(1996){Matthews}, {Ghez}, {Weinberger}, \&
  {Neugebauer}}]{matthews96}
{Matthews}, K., {Ghez}, A.~M., {Weinberger}, A.~J., \& {Neugebauer}, G. 1996,
  \pasp, 108, 615+

\bibitem[{{Monnier} {et~al.}(2006){Monnier}, {Berger}, {Millan-Gabet}, {Traub},
  {Schloerb}, {Pedretti}, {Benisty}, {Carleton}, {Haguenauer}, {Kern},
  {Labeye}, {Lacasse}, {Malbet}, {Perraut}, {Pearlman}, \&
  {Zhao}}]{monnier2006}
{Monnier}, J.~D., et al.  2006, \apj, 647, 444

\bibitem[{{Monnier} {et~al.}(2007){Monnier}, {Tuthill}, {Danchi}, {Murphy}, \&
  {Harries}}]{monnier2007a}
{Monnier}, J.~D., et~al.  2007, \apj, 655, 1033

\bibitem[{{Monnier} {et~al.}(2004){Monnier}, {Tuthill}, {Ireland}, {Cohen}, \&
  {Tannirkulam}}]{monnier_aas2004}
{Monnier}, J.~D., {Tuthill}, P.~G., {Ireland}, M.~J., {Cohen}, R., \&
  {Tannirkulam}, A. 2004, in Bulletin of the AAS,
  Vol.~36, 1367--+

\bibitem[{{Palla} \& {Stahler}(1993)}]{palla1993}
{Palla}, F. \& {Stahler}, S.~W. 1993, \apj, 418, 414

\bibitem[{{Pauls} {et~al.}(2005){Pauls}, {Young}, {Cotton}, \&
  {Monnier}}]{pauls2005}
{Pauls}, T.~A., {Young}, J.~S., {Cotton}, W.~D., \& {Monnier}, J.~D. 2005,
  \pasp, 117, 1255

\bibitem[{{Perrin} {et~al.}(1998){Perrin}, {Coude Du Foresto}, {Ridgway},
  {Mariotti}, {Traub}, {Carleton}, \& {Lacasse}}]{perrin1998}
{Perrin}, G., et~al.
{Coude Du Foresto}, V., {Ridgway}, S.~T., {Mariotti}, J.-M., 1998, \aap, 331, 619

\bibitem[{{Rajagopal} {et~al.}(2007){Rajagopal}, {Menut}, {Wallace}, {Danchi},
  {Chesneau}, {Lopez}, {Monnier}, {Ireland}, \& {Tuthill}}]{rajagopal2007}
{Rajagopal}, J., et al.  2007, \apj, 671, 2017


\bibitem[{{Schneider} {et~al.}(1999){Schneider}, {Smith}, {Becklin}, {Koerner},
  {Meier}, {Hines}, {Lowrance}, {Terrile}, {Thompson}, \&
  {Rieke}}]{schneider1999}
{Schneider}, G., et~al. 1999, \apjl, 513, L127

\bibitem[{{Smith} {et~al.}(2005){Smith}, {Balega}, {Duschl}, {Hofmann},
  {Lachaume}, {Preibisch}, {Schertl}, \& {Weigelt}}]{smith2005}
{Smith}, K.~W., et~al.  2005, \aap, 431, 307

\bibitem[{{Strom} \& {Strom}(1994)}]{strom1994}
{Strom}, K.~M. \& {Strom}, S.~E. 1994, \apj, 424, 237

\bibitem[{{Tuthill} {et~al.}(2000){Tuthill}, {Monnier}, {Danchi}, {Wishnow}, \&
  {Haniff}}]{tuthill2000b}
{Tuthill}, P.~G., et~al.  2000, \pasp, 112, 555

\bibitem[{{Tuthill} {et~al.}(2008){Tuthill}, {Monnier}, {Lawrance}, {Danchi},
  {Owocki}, \& {Gayley}}]{tuthill2008a}
{Tuthill}, P.~G., et~al.  2008, \apj, 675, 698

\bibitem[{{Weiner} {et~al.}(2006){Weiner}, {Tatebe}, {Hale}, {Townes},
  {Monnier}, {Ireland}, {Tuthill}, {Cohen}, {Barry}, {Rajagopal}, \&
  {Danchi}}]{weiner2006}
{Weiner}, J., et~al.  2006, \apj, 636, 1067


\end{thebibliography}

\clearpage

\clearpage


\clearpage

\end{document}